\documentstyle[11pt,./newpasp,./psfig,twoside]{article}
\def\edcomment#1{\iffalse\marginpar{\raggedright\sl#1\/}\else\relax\fi}
\reversemarginpar

\newcommand{\Wo}{\mbox{${\rm W}_0$}}
\newcommand{\msun}{\mbox{${\rm M}_\odot$}}

\newcommand{\SeBa}{\mbox{${\sf SeBa}$}}

\newcommand{\nbody}{\mbox{{{\em N}-body}}}
\newcommand{\thm}{\mbox{${t_{\rm hm}}$}}

\newcommand{\trlx}{\mbox{${t_{\rm rlx}}$}}

\newcommand{\mgal}{\mbox{${M_{\rm Gal}}$}}

\newcommand{\rcore}{\mbox{${r_{\rm core}}$}}

\newcommand{\rhm}{\mbox{${r_{\rm hm}}$}}
\newcommand{\rgc}{\mbox{${r_{\rm GC}}$}}

\newcommand{\rtide}{\mbox{${r_{\rm tide}}$}}

\newcommand{\rLf}{\mbox{${r_{\rm L1}}$}}

\def\unit#1{{\mbox{[{\rm #1}]}}}
\def\apgt{\ {\raise-.5ex\hbox{$\buildrel>\over\sim$}}\ }
\def\aplt{\ {\raise-.5ex\hbox{$\buildrel<\over\sim$}}\ }

\begin{document}
\title{Is the Galactic center populated with young star clusters?}
\author{Simon Portegies Zwart}
\affil{Dept.\ of Astronomy,
		  Boston University,
		  725 Commonwealth Ave.,
		  Boston, MA 02215, USA\\ Hubble Fellow}


\begin{abstract}
We study the evolution and observability of young and compact star
clusters near the Galactic center, such as the Arches cluster and the
Quintuplet. The star clusters are modeled with a combination of
techniques; using direct \nbody\, integration to calculate the motions
of all stars and detailed stellar and binary evolution to follow the
evolution of the stars.
The modeled star clusters dissolve within 10 to 60 million years in
the tidal field of the Galaxy. The projected stellar density in the
modeled clusters drops within 5\% to 70\% of the lifetime to a level
comparable to the projected background density towards the Galactic
center. And it will be very hard to distinguished these clusters at
later age among the background stars.  This effect is more severe for
clusters at larger distance from the Galactic center but in projection
at the same distance. Based on these arguments we conclude that the
Galactic center easily hides 10 to 40 clusters with characteristics
similar to the Arches and the Quintuplet cluster.
\end{abstract}

\section{Introduction}
A number of compact and young star clusters have been observed within
the inner few ten parsec from the Galactic center.  Most
noticeable are the Arches cluster (Object 17, Nagata et al.\,
1995) and the Quintuplet cluster
(AFGL\,2004, Nagata et al.\, 1990; Okuda et al.\,
1990). But it
is not excluded that others exist as these clusters are well hidden
behind thick layers of absorbing material. The Arches and the
Quintuplet clusters form the galactic counterparts of NGC\,2070 (or
R\,136); the central star cluster in the 30\,Doradus region in the
Large Magellanic Could (Massey \& Hunter
1998).  The structural parameters of these
clusters, size, mass and density profile are quite similar as are
their ages.  The Arches and the Quintuplet clusters are at a projected
distance of $\aplt 50$\,pc from the Galactic center.  Their evolution
is therefore dramatically affected by the presence of the tidal field
of the Galactic bulge and inner disc.

In this paper we report the results of \nbody\ simulations of young
and compact star clusters, such as the Arches cluster and the
Quintuplet cluster in the vicinity of the Galactic bulge.  These
cluster are particularly interesting because a strong coupling between
stellar evolution, stellar dynamics and the tidal field of the Galaxy
may exist.  In addition to this, excellent observational data are
available.  Many unusually bright and massive stars are present in
both clusters which, due to the high central density of $10^5$ to
$10^6$ stars pc$^{-3}$ are likely to interact strongly with each
other.

A number of intriguing question about these clusters makes it worth to
model them in great detail, these are 1.) are these clusters the
progenitors of globular star clusters, 2.) what is their contribution
to the star formation rate in the Galaxy, 3.) are their mass functions
intrinsically flat as has been suggested by observations, 4.) how far
are these clusters really from the Galactic center and 5.) how many
are still hidden, waiting to be discovered.  I will address the latter
two conundrums in this paper and a more detailed paper is in
preparation (Portegies Zwart, Makino, McMillan \& Hut, 2000b).

\section{The model}
We study the evolution of the Arches and Quintuplet cluster by
integrating the equations of motion of all stars and at the same time
we account for the evolution of the stars and binaries. The adopted
\nbody\ integration algorithm, evolution of stars and binaries, and
the interface between the dynamical calculations and the stellar
evolution are described extensively by Portegies Zwart et al.\, (2000a,
see also {\tt http://www.sns.ias.edu/$\sim$starlab}).

The \nbody\ portion of the simulations is carried out using {\tt
kira}, operating within the Starlab software environment
(Portegies Zwart et al.\ 1998). Time integration of
stellar orbits is accomplished using a fourth-order Hermite scheme
(Makino \& Aarseth 1992). {\tt Kira} also incorporates
block timesteps (McMillan 1986a; 1986b; Makino
1991) special treatment of close two-body and
multiple encounters of arbitrary complexity, and a robust treatment of
stellar and binary evolution and stellar collisions.  The
special-purpose GRAPE-4 (Makino et al.\
1997) system is used to accelerate the
computation of gravitational forces between stars. 

The evolution of stars and binaries are carried out by \SeBa\, (see
Portegies Zwart \& Verbunt, 1996, Sect.\, 2.1) the
binary evolution package which is combined in the starlab software
toolset.  The treatment of collisions and mass loss in the
main-sequence stage for massive stars are described by Portegies Zwart
et al.\, (1998; 1999; 2000a).

\section{Initial conditions}

The observed parameters for the Arches and the Quintuplet clusters are
presented in Tab.\,1. These clusters have masses of
about $\sim 10^4$\,\msun\, and are extremely compact $\rhm\aplt 1$\,pc
(Figer, McLean \& Morris 1999). The
projected distance from the Arches cluster to the Galactic center is
about 34\,pc, the Quintuplet cluster is somewhat farther away.  If the
third component of the projected distance to the Galactic center is
zero then the observed distance equals the real distance.

\begin{table*}[ht]
\caption[]{Observed parameters for some of these clusters.  The
columns give the cluster name, reference, age, mass, distance to the
Galactic center, the tidal radius and the half mass radius.  The last
column gives the density within the half mass radius.
}
\begin{flushleft}
\begin{tabular}{ll|rrrrrr} \hline
Name 	  &ref& Age  &   M     & \rgc & \rtide & \rhm   
				& $\rho_{\rm hm}$ \\ 
          &&[Myr]& [$10^3$\,\msun] & \multicolumn{3}{c}{[pc]}   &
			 [$10^5$\,\msun/pc$^2$] \\ \hline
Arches    &a& 1--2 & 12--50   & 30   & 1       &  0.2    & 0.3 \\ 
Quintuplet&b& 3--5 & 10--16   & 50   & 1       &$\aplt 1$& 0.2 \\
\hline
\end{tabular} \\
References:
a) Brandl et al.\, (1996);
   Campbell et al.\, (1992);
   Massey \& Hunter (1998).
b) Figer et al.\, (1999);
\end{flushleft}
\label{Tab:observed} 
\end{table*}

Our calculations start with 12k (12288) stars at zero age.  We assign
masses to stars between 0.1\,\msun\ and 100\,\msun\ from the mass
function suggested for the Solar neighborhood by Scalo
(1986). The median mass of this mass function is about
0.3\,\msun, and the mean mass $\langle m \rangle \simeq 0.6\,\msun$.
For the models with 12k stars this results in a total cluster mass of
$\sim 7\,500$\,\msun.  Initially all stars are single, but binaries
may form via three body encounters in which one star carries away the
excess energy and angular momentum.  We adopt two distances from the
Galactic center, 34\,pc and 150\,pc.  The initial density profile and
velocity dispersion for the models are taken from a Heggie \& Ramamani
(1995) model with $\Wo = 4$. At birth the clusters are
assumed to perfectly fill the zero velocity surface in the tidal field
of the Galaxy.  An overview of the initial conditions for the computed
models is summarized in Tab.\,2.

\begin{table*}[ht]
\caption[]{ Overview of initial conditions for the calculations.  Each
row stars with the model name, the distance to the Galactic center and
the initial King model \Wo, then the initial relaxation time, the half
mass crossing time and the core radius, half mass radius and the tidal
radius. The last two columns give the the time of core collapse and
the time the cluster mass drops below 5\% in the initial mass (or
about 375\,\msun).  }\begin{flushleft}
\begin{tabular}{lrl|rrrll|rrr} \hline
Model  &$r_{\rm GC}$
            &\Wo&\trlx& \thm&\rcore&\rhm & \rtide     &$t_{cc}$ 
						      &$t_{\rm end}$ \\
       &[pc]&   &[Myr]&[Kyr]&\multicolumn{3}{c}{[pc]} &
			    &\multicolumn{2}{c}{[Myr]} \\ 
\hline
R34W1  & 34 & 1&  5.5 &  43 & 0.12 & 0.17& 0.7  &1.9&  9.3 \\ 
R34W4  & 34 & 4&  3.3 &  26 & 0.08 & 0.12& 0.7  &0.9& 12.4 \\ 
R150W4 &150 & 4& 12.9 & 104 & 0.22 & 0.30& 1.9  &3.2& 58.6 \\ 
\hline     		       
\end{tabular}
\end{flushleft}
\label{Tab:N12kinit} \end{table*}

The tidal field is characterized by the Oort (1927)
constants $A$ and $B$ and the local stellar density.  The mass within
the clusters' orbit at a distant \rgc\, from the Galactic center is
calculated with (Mezger et al.\, 1999)
\begin{equation}
	\mgal(\rgc) = 4.25\, 10^6 \left({\rgc \over \unit{pc}}
		     	           \right)^{1.2}           \;\;[\msun].
\label{Eq:Mgal}\end{equation}
And with this we can derive the appropriate Oort constants.

Once the tidal field, the mass of the cluster and its density profile
are selected the \nbody\, system is fully determined.  The total mass
of the stellar system determines the unit of mass in the \nbody\,
system, the distance to the first Lagrangian point \rLf\, in the tidal
field of the Galaxy sets the distance unit and the velocity dispersion
together with the size of the stellar system sets the time scale.  The
evolution of the cluster is subsequently followed using the direct
\nbody\, integration including stellar and binary evolution and the
tidal field of the Galaxy (see Portegies Zwart et al.\, 2000b).

For economic reasons not all stars are kept in the \nbody\, system,
but stars are removed when they are 3\rLf\, from the center of the
star cluster.

\section{Results}
\begin{figure}[htbp!]
\hspace*{1.cm}
\psfig{figure=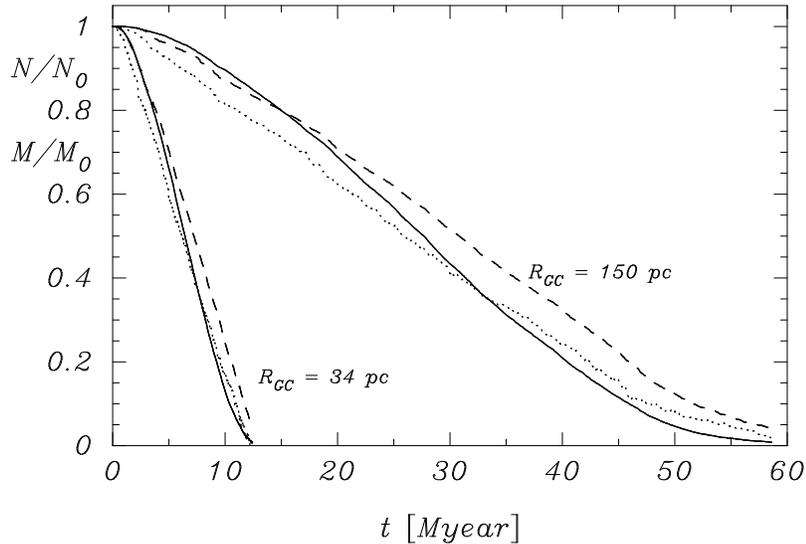,width=12.cm,angle=-90}
\caption[]{Evolution of the total mass $M$ (dashes), the bound mass
(dots) and number of stars $N$ (solids) for the models with $\Wo=4$ at
$\rgc=34$\,pc (left) and $\rgc=150$\,pc (right).  Cluster farther away
from the Galactic center live longer. }
\label{fig:SPZ_W147N12k_TM}
\end{figure}

Figure\,1 shows the evolution of the mass and
number of stars of two models from Table.\,2.  Star
clusters which are located further away from the galactic center live
considerably longer than the closer clusters. Estimating the cluster
lifetime naively via the initial relaxation time would lead to an age
of 48\,Myear ($\equiv 12.4 {\rm Myear} \times 12.9/3.3$) for model
R150W4. The weaker tidal field at a distance of 150\,pc from the
Galactic center, however, tend to extend the cluster lifetime to about
59\,Myear, indicating that the strength of the tidal field has a
stronger effect on the cluster lifetime than mass loss from stellar
evolution.

The number of stars (solid) in the models (see
Fig.\,1) decreases more quickly than the total
mass (dashes and dots). The mean mass of the stars within the cluster
potential therefore increases gradually with time.  Near $t=7$\,Myear
for model R34W4 and $t=32$\,Myear for model R150W4 the number of stars
drops below the total cluster mass, indicating that the mean mass
exceeds 1\,\msun.  At these moments both models have lost $\sim 65$\%
of their initial mass.


\section{Discussion}

Although the Arches and Quintuplet clusters are very compact, it may
still be hard to notice them near the Galactic center.  The local
stellar density is high and we can only see these clusters in
projection onto the background.

The projected density near the galactic center can be calculated by
differentiating Eq.\,1 with respect to the distance along
line of sight. Portegies Zwart et al.\,(2000b) perform this calculation
numerically and arrive at a projected density of about
3000\,\msun\,pc$^{-3}$. 

\begin{figure}[htbp!]
\hspace*{1.cm}
\psfig{figure=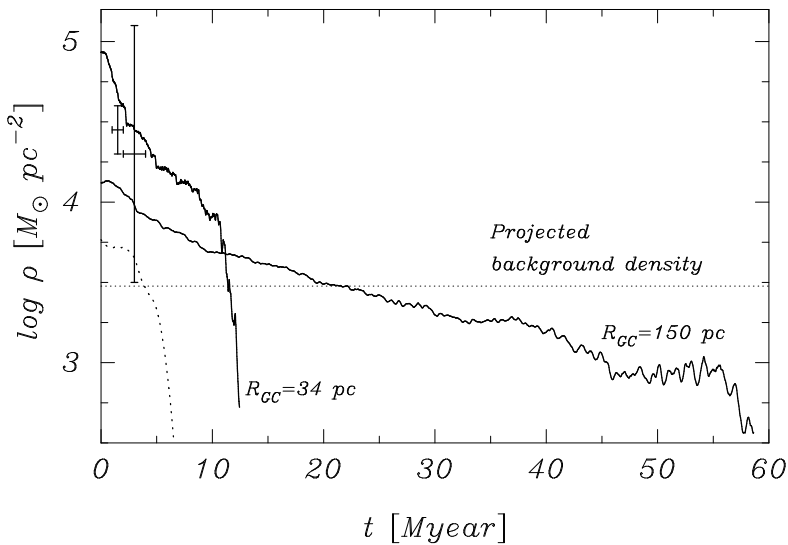,width=12.cm,angle=-90}
\caption[]{ Evolution of the projected density within the projected
half mass radius for the models at $\rgc = 34$\,pc (left) and at $\rgc
= 150$\,pc (right). The dashed line gives the evolution of the half
mass density for a model with $\Wo=1$ at $\rgc = 34$\,pc.  The
horizontal dotted line gives the projected background density in the
direction of a projected distance of 34\,pc from the Galactic center.
The two error bars give the observed projected density of the Arches
and the Quintuplet clusters.}
\label{fig:SPZ_W17R34_rho_phm}
\end{figure}

Figure\,2 shows the evolution of the
density within the projected half mass radius for models R150W1
(dotes), R34W4 (left solid) and R150W4 (right solid). The two error
bars give the projected half mass densities for the observed clusters:
Arches (left) and Quintuplet (right). The horizontal dotted line gives
the surface density at a projected distance of 34\,pc from the
Galactic center.

The projected densities of the observed clusters are about an order of
magnitude higher than the background. Clusters with a lower background
density may remain unnoticed among the background stars, as observers
may have difficulty to distinguish the cluster from the background.

The projected density of model R34W1 barely exceeds the background and
such a cluster easily remains unnoticed throughout its lifetime.  The
two models which started more concentrated R34W4 and R150W4 have
projected densities well above the background, at least initially.
The cluster farther away from the Galactic center has a lower density
because it is more extended; its tidal radius is larger. After the
first few million years this cluster may be hard to notice among the
dense stellar background. We adopt a minimum contrast about three
times the projected background density, i.e.: $10^4$\,\msun\,pc$^{-2}$
required for distinguishing a star cluster among the background.  In
that case the cluster at a distance of 150\,pc would be visible for
only about 3\,million years ($\sim 5$\% of its lifetime), and almost
9\,million years ($\sim 70$\% of its lifetime) for the cluster at a
distance of 34\,pc. Although the cluster far away from the Galactic
center (model R150W4) lives much longer, it only remains visible for a
small fraction of its lifetime.

Expressed in fraction of their lifetime the two models are observable
for about 5\% to 70\% of the time for the clusters at a distance of
150\,pc and 34\,pc, respectively. We therefore expect that a large
population of clusters with characteristics similar to the Arches and
Quintuplet may still be hidden in the direction of the Galactic
center. Based on our calculations we estimate that more than 40
clusters remain to be found within a projected distance of 50\,pc from
the Galactic center. Most of these will be older than the Arches and
Quintuplet clusters but not exceeding $10^8$ years, as they dissolve
on a shorter time scale.

\section{Conclusion}
We studied the evolution of the two young and compact star clusters
near the Galactic center, the Arches cluster and the Quintuplet. Based
on the background stellar density towards the Galactic center and the
projected density of the modeled and observed clusters we conclude
that the number of such clusters must be far greater than
observed. Most of these clusters remain hidden in the stellar
background and they are observable only in the first few million years
of their existence, when they are still very compact.  Within a
projected distance of 50\,pc from the Galactic center between 10 and
40 clusters with characterisitcs similar to the Arches cluster and the
Quintuplet may be hidden. These hidden clusters are likely to be
somewhat older and less compact than those already found.

\acknowledgements 

I thank Piet Hut, Ken Janes, Jun Makino and Steve McMillan for
discussions.  I am grateful to Drexel University and Tokyo University
for the use of their GRAPE hardware.  This work was supported by NASA
through Hubble Fellowship grant HF-01112.01-98A awarded (to SPZ) by
the Space Telescope Science Institute, which is operated by the
Association of Universities for Research in Astronomy.



\end{document}